\documentclass[aps,final,notitlepage,oneside,nobibnotes,nofootinbib,superscriptaddress,centertags,showpacs]{revtex4-1}
\usepackage{pxfonts}
\begin{document}
\title{Phenomenology of particle creation in Weyl geometry}
\author{Victor Aleksandrovich Berezin}\thanks{e-mail: berezin@inr.ac.ru}
\affiliation{Institute for Nuclear Research of the Russian Academy of Sciences
	60th October Anniversary Prospect 7a, 117312 Moscow, Russia}
\author{Vyacheslav Ivanovich Dokuchaev}\thanks{e-mail: dokuchaev@inr.ac.ru}
\affiliation{Institute for Nuclear Research of the Russian Academy of Sciences
60th October Anniversary Prospect 7a, 117312 Moscow, Russia}
\date{\today}
\begin{abstract}
Short review of the Weyl geometry is given. To describe the phenomenological particle creation we suggest the modified perfect fluid model taking into account the back reaction on the geometry of both the already created particles and the very process of their creation. It is found that the relation for particle creation is conformal invariant. This requires the creation law consisting of the source terms as the Weyl  Lagrangian plus two quite new terms depending of the particle number density.
\end{abstract}
%\keywords{gravitation, Weyl geometry, General Relativity, quadratic gravity, cosmology}
\maketitle

\section{Introduction}

One of the most fundamental problems of the theoretical physics is the origin of gravity. A. D. Sakharov in 1966 \cite{Sakharov66} formulated the very fruitful ideas on the origin of gravity and on the possible initial state of the Universe. These ideas were supported and generalized later {besides the other scientists} by A. Vilenkin \cite{Vilenkin}, R. Penrose \cite{Penrose} and G. 't Hooft \cite{Hooft}. 

Nowadays there are numerous propositions for the gravity theories for the modification and generalization of the Einstein general relativity. Hermann Weyl  \cite{Weyl} in 1918 year developed very profound conformal gravity theory, which historically was the first modification of the Einstein general relativity. The local conformal invariance seems to be a good candidate to become  the fundamental symmetry of the Nature. 

In this paper we shortly review the basic features Weyl geometry. To describe the phenomenological particle creation we suggest the modified perfect fluid model taking into account the back reaction on the geometry of both the already created particles and the very process of their creation. It is found that the relation for particle creation is conformal invariant. This requires the creation law consisting of the source terms as the Weyl  Lagrangian plus two quite new terms depending of the particle number density. We call this mechanism of particle creation in the Weyl geometry as ``{\it creating  the universe from nothing}''{}.

\section{Weyl differential geometry}

Any differential geometry is uniquely determined by the metric tensor $g_{\mu\nu}$ and connections $\Gamma^\lambda_{\mu\nu}(x)$. The metric tensor defines the interval $ds$ between nearby points,
\begin{equation} \label{ds2}
	ds^2=g_{\mu\nu}dx^\mu dx^\nu,
\end{equation}
and the connections  $\Gamma^\lambda_{\mu\nu}(x)$ are needed in construction of the parallel transport of and covariant derivatives, $\nabla$, for vectors. Namely, the  covariant derivative of some vector $l^\mu$ is defined by 
\begin{equation} \label{nabla}
	\nabla_{\lambda} l^\mu= l^\mu_{\;,\lambda} +\Gamma_{\lambda\nu}^\mu l^\nu,
\end{equation}
while for its parallel transfer one has
\begin{equation} \label{nabla2}
	\nabla_{\lambda} l^\mu dx^\lambda=0.
\end{equation}
Here comma ``,'' stands for the usual partial derivative.

By definition, the covariant derivative of a scalar field $\varphi$ is just its partial derivative, $\nabla_\mu \varphi=\varphi_{,\mu}$. Moreover, any derivative is a linear operator obeying the Leibniz rule for the products. Immediate consequence of this is formula for the covariant derivatives for the covariant vectors, $l_\mu$,
\begin{equation} \label{nabla3}
	\nabla_\mu l_\nu= l_{\nu,\mu} -\Gamma_{\mu\nu}^\lambda l_\lambda,
\end{equation}
and for tensors, for example,
\begin{eqnarray} \label{tensot}
	\nabla_\mu A^{\nu\lambda}&=& A^{\nu\lambda}_{\mu} +\Gamma_{\mu\sigma}^\nu A^{\sigma\lambda} +\Gamma_{\mu\sigma}^\lambda A^{\nu\sigma}\!, \\
	\nabla_\mu A^\nu_\lambda \ &=& A^\nu_{\lambda,\mu} +\Gamma_{\mu\sigma}^\nu A^\sigma_\lambda -\Gamma_{\mu\lambda}^\sigma A^\nu_\sigma,
\end{eqnarray}
and so on.

Covariant derivative procedure transforms scalar fields into the vector fields, vector fields into the tensor fields of the second rank and so on.

Since the partial derivative of any vector is nt a tensor, the connections $\Gamma_{\mu\nu}^\lambda$ is not a tenor as well. But, it is possible to construct a tensor (different from $g_{\mu\nu}$), using connections and its first derivatives. This is the curvature tensor 
\begin{equation} \label{curv}
	R^{\mu}_{\phantom{\mu}\nu\lambda\sigma}=\frac{\partial \Gamma^\mu_{\nu\sigma}}{\partial x^\lambda}-\frac{\partial \Gamma^\mu_{\nu\lambda}}{\partial x^\sigma}+\Gamma^\mu_{\varkappa\lambda}\Gamma^\varkappa_{\nu\sigma}-\Gamma^\mu_{\varkappa\sigma}\Gamma^\varkappa_{\nu\lambda},
\end{equation}
which reflects the structure of spacetime in the neighborhood of given point.
By contraction one obtains the Ricci tensor 
\begin{equation} \label{Ricci}
	R_{\mu\nu}=R^{\lambda}_{\phantom{\mu}\mu\lambda\nu} 
\end{equation}
and the curvature scalar
\begin{equation} \label{R}
	R=R_\lambda^\lambda.
\end{equation}
The raising and lowering of indices is performed by contracting with the metric tensor, $g_{\mu\nu}$, or its inverse, $g^{\mu\nu}$, $g^{\mu\nu}g_{\nu\lambda}=\delta^\mu_\lambda$, $\delta^\mu_\lambda$ is the Kronecker symbol (unit tensor).

It appears that connections can be computed if one knows three tensors, the metric tensor $g_{\mu\nu}$ (and its inverse $g^{\nu\lambda}$), the torsion tensor
$S^\lambda_{\phantom{\mu}\mu\nu}$ and the nonmetricity tensor $Q_{\lambda\mu\nu}$, where
\begin{equation} \label{S}
	S^\lambda_{\phantom{\mu}\mu\nu}=\Gamma^\lambda_{\mu\nu} -\Gamma^\lambda_{\mu\nu}, 
\end{equation}
\begin{equation} \label{S2}
	Q_{\mu\nu\lambda}= \nabla_\mu g_{\nu\lambda}.
\end{equation}
Evidently, 
\begin{equation} \label{S3}
	S^\lambda_{\phantom{\mu}\mu\nu} =-S^\lambda_{\phantom{\mu}\nu\mu}
\end{equation}
and 
\begin{equation} \label{S4}
	Q_{\mu\nu\lambda}= Q_{\mu\lambda\nu}.
\end{equation}

The corresponding relation for the connections is the following 
\begin{equation} \label{Gamma}
	\Gamma^\lambda_{\mu\nu}=C^\lambda_{\phantom{\mu}\mu\nu} + K^\lambda_{\phantom{\mu}\mu\nu} +L^\lambda_{\phantom{\mu}\mu\nu},
\end{equation}
where $C^\lambda_{\phantom{\mu}\mu\nu}$ are the famous Christoffel symbols,
\begin{equation} \label{C}
	C^\lambda_{\phantom{\mu}\mu\nu}=\frac{1}{2}g^{\lambda\kappa} (g_{\kappa\mu,\nu}+g_{\kappa\nu,\mu}-g_{\mu\nu,\kappa})
\end{equation}
and
\begin{equation} \label{K}
	K^\lambda_{\phantom{\mu}\mu\nu}=\frac{1}{2} (S^\lambda_{\phantom{\mu}\mu\nu}-S^{\phantom{\mu}\lambda}_{\mu\phantom{\mu}\nu} -S^{\phantom{\mu}\lambda}_{\nu\phantom{\mu}\mu}),
\end{equation}
\begin{equation} \label{L}
	L^\lambda_{\phantom{\mu}\mu\nu}=\frac{1}{2} (Q^\lambda_{\phantom{\mu}\mu\nu}-Q^{\phantom{\mu}\lambda}_{\mu\phantom{\mu}\nu} -Q^{\phantom{\mu}\lambda}_{\nu\phantom{\mu}\mu}).
\end{equation}
By using the Christoffel symbols as the connections, we can construct another covariant derivative, the metric one, different from $\nabla_\mu$, which will be denoted by a semicolon ``$;$''. 

Within this scheme the Riemannian geometry is the simplest one. Namely, 
\begin{equation} \label{R3a}
	S^\lambda_{\phantom{\mu}\mu\nu}=0,
\end{equation}
\begin{equation} \label{R3b}
	Q_{\lambda\mu\nu}=0.
\end{equation}
Moreover, apart from the obvious relation 
\begin{equation} \label{R4b}
	R^\mu_{\nu\lambda\sigma}=-R^\mu_{\nu\sigma\lambda},
\end{equation}
the curvature tensor $R_{\mu\nu\lambda\sigma}$ obeys some additional algebraic identities,
\begin{equation} \label{R4}
	R_{\mu\nu\lambda\sigma}=-R_{\nu\mu\lambda\sigma}= R_{\lambda\sigma\mu\nu},
\end{equation}
and the differential Bianchi identity
\begin{equation} \label{Bianchi}
	R^\mu_{\phantom{\mu}\nu\lambda\sigma;\kappa}  +R^\mu_{\phantom{\mu}\nu\kappa\lambda;\sigma} + R^\mu_{\phantom{\mu}\nu\sigma\kappa;\lambda}=0.
\end{equation}
In addition, Ricci tensor is symmetric, 
\begin{equation} \label{Ricci2}
	R_{\mu\nu}=R_{\nu\mu}.
\end{equation}

The Weyl geometry belongs to the next level of sophistication. The torsion tensor is still zero, 
\begin{equation} \label{Q20}
	S^\lambda_{\phantom{\mu}\mu\nu}=0, 
\end{equation}
but the nonmetricity tensor is not:
\begin{equation} \label{Q2}
	Q_{\mu\nu\lambda}=A_\mu g_{\nu\lambda},
\end{equation}
where $A_\mu$ is called the ``Weyl vector''. The connections becomes now
\begin{equation} \label{Wb}
	\Gamma^\lambda_{\mu\nu}=C^\lambda_{\mu\nu}+W^\lambda_{\mu\nu},
\end{equation}
where
\begin{equation} \label{W}
	W^\lambda_{\mu\nu}=-\frac{1}{2}(A_\mu \delta^\lambda_\nu+ A_\nu \delta^\lambda_\mu -A^\lambda g_{\mu\nu}),
\end{equation}
and $\delta^\lambda_\nu$ is the Kronecker symbol. 

The algebraic identities for the curvature tensor $R_{\mu\nu\lambda\sigma}$ are not so trivial. One has 
\begin{equation} \label{R4}
	R_{\mu\nu\lambda\sigma}+R_{\nu\mu\lambda\sigma}=2F_{\lambda\sigma}g_{\mu\nu},
\end{equation}
\begin{equation} \label{R3}
	R_{\mu\nu\lambda\sigma}-R_{\lambda\sigma\mu\nu}= \frac{1}{2}\left(F_{\mu\sigma}g_{\nu\lambda} -F_{\mu\lambda}g_{\nu\sigma} 
	+F_{\nu\lambda}g_{\mu\sigma} - F_{\nu\sigma}g_{\mu\lambda} +F_{\mu\nu}g_{\lambda\sigma} -F_{\lambda\sigma}g_{\mu\nu}\right), 
\end{equation}
where
\begin{equation} \label{Stress}
	F_{\mu\nu}=\nabla_\mu A_\nu - \nabla_\nu A_\mu =A_{\nu;\mu} - A_{\mu;\nu} =A_{\nu,\mu} - A_{\mu,\nu}.
\end{equation}
In spite of this, the cyclic identity transformation remains untouched, as well as Bianchi identity (surely with replacing the metric covariant derivatives to the Weyl's ones). 

Also the Ricci tensor is no more symmetric,  
\begin{equation} \label{Ricci5}
	R_{\nu\sigma}-R_{\sigma\nu}= F_{\nu\sigma}. 
\end{equation}

\section{Conformal invariance and Weyl gravity}

In 1919 Hermann Weyl, trying to build the unified geometric theory of electromagnetic and gravitational interactions, discovered that Maxwell equations are invariant under the local conformal transformations and claimed that gravitation should be invariant as well.

By definition, the local conformal transformation converts the interval $ds$ between the nearby points into $\hat ds$ by relation 
\begin{equation} \label{local}
	ds^2=g_{\mu\nu}dx^\mu dx^\nu = \Omega^2(x)d\hat s^2=\Omega^2(x)\hat g_{\mu\nu}(x)dx^\mu dx^\nu,
\end{equation}
$\Omega(x)$ is called the ``conformal factor''. Note that coordinate system remains untouched. 

In order to achieve his goal, Weyl postulated that the connections $\Gamma_{\mu\nu}$ must be conformal invariant, i.\,e., 
\begin{equation} \label{local2}
	\Gamma_{\mu\nu}^\lambda =\hat\Gamma_{\mu\nu}^\lambda.
\end{equation}
This means that if any vector undergoes the parallel transfer, the same should be happen in the conformally transformed metric. Since, for the Christoffel symbols one has
\begin{eqnarray} \label{hatC}
	C^\lambda_{\mu\nu}&=&\frac{1}{2\Omega^2}\hat g^{\lambda\sigma}\left((\Omega^2\hat g_{\sigma\mu})_{,\nu} +(\Omega^2\hat g_{\sigma\nu})_{,\mu} -(\Omega^2\hat g_{\mu\nu})_{,\sigma}\right)
	\nonumber \\ 
	&=&\hat C^\lambda_{\mu\nu} +\left(\frac{\Omega_{,\mu}}{\Omega}\delta^\lambda_\nu +\frac{\Omega_{,\nu}}{\Omega}\delta^\lambda_\mu -\frac{\Omega_{,\kappa}}{\Omega}\hat g^{\lambda\kappa}\hat g_{\mu\nu}\right)\!,
\end{eqnarray}
then the required conformal invariance of connections implies
\begin{equation} \label{Ahat}
	A_\mu=\hat A_\mu+2\frac{\Omega_{,\mu}}{\Omega}.
\end{equation}
The Weyl vector $A_\mu$ turned out the gauge field, like the electromagnetic vector.

The curvature tensor $R^\mu_{\phantom{1}\nu\lambda\sigma}$ is constructed solely of connections. Therefore
\begin{equation} \label{R6}
	R^\mu_{\phantom{1}\nu\lambda\sigma}=\hat R^\mu_{\phantom{1}\nu\lambda\sigma}.
\end{equation}
The same is true for  the Ricci tensor 
\begin{equation} \label{Ricci6}
	R_{\mu\nu}=\hat R_{\mu\nu}.
\end{equation}
Evidently,
\begin{equation} \label{F6}
	F_{\mu\nu}=\hat F_{\mu\nu}.
\end{equation}

In analogy with classical electrodynamics, H. Weyl postulated the following transparently conformal invariant action
\begin{equation} \label{SW}
	S_{\rm W} =\int\!{\cal L_{\rm W}}\sqrt{-g}\,d^4x,
\end{equation}
with
\begin{equation} \label{LW}
	{\cal L_{\rm W}}=\alpha_1  R_{\mu\nu\lambda\sigma}R^{\mu\nu\lambda\sigma}
+\alpha_2R_{\mu\nu}R^{\mu\nu} +\alpha_3R^2 +\alpha_4F_{\mu\nu}F^{\mu\nu}.
\end{equation}
We called this ``the Weyl gravity'' \cite{bde19,bde20,bdes,bd23}. 

Note again that the geometric dynamical variables are the metric tensor $g_{\mu\nu}(x)$ and the Weyl Vector $A_\mu(x)$.

The total action consists of two parts, the gravitational one, $S_{\rm W}$, and the action $S_{\rm m}$ for the matter parts,
\begin{equation} \label{total}
	S_{\rm tot}=S_{\rm W}+S_{\rm m},
\end{equation}
\begin{equation} \label{total2}
	S_{\rm m}= \int\!{\cal L_{\rm m}}\sqrt{-g}\,d^4x.
\end{equation}

Being zero on the field equations, the variation of total action is automatically conformal  invariant there. But, the Weyl's geometrical action is constructed to be conformal invariant --- so does its variation.

Consequently, the variation of action for matter fields has to be conformal invariant, as well. Let us introduce the definition 
\begin{equation} \label{R3}
		\delta S_{\rm m}\stackrel{\mathrm{def}}{=} - \frac{1}{2}\!\int T^{\mu\nu}(\delta g_{\mu\nu})\sqrt{-g}\,d^4x -\!\int G^\mu(\delta A_\mu)\sqrt{-g}\,d^4x
	+\!\int\frac{\delta \cal L_{\rm  W}}{\delta\Psi} (\delta \Psi)\sqrt{-g}\,d^4x,
\end{equation}
where $T^{\mu\nu}$ is the energy-momentum tensor, $G^\mu$ can be called ``the Weyl current'', and $\Psi$ is the collective dynamical variable, describing the matter field, the matter motion being determined by the Lagrange equation $\delta S/\delta\Psi=0$.

We already know that 
\begin{equation} \label{deltag}
	\delta g_{\mu\nu}=2\Omega(\delta\Omega)\hat g_{\mu\nu} =2g_{\mu\nu}\!\left(\frac{\delta\Omega}{\Omega}\right)\!.
\end{equation}
\begin{equation} \label{deltag}
	\delta A_\mu=2\delta\left(\!\frac{\Omega_{,\mu}}{\Omega}\!\right) =2\left(\delta(\log\Omega)\right)_{,\mu}
	=2\left(\!\frac{\delta\Omega_{,\mu}}{\Omega}\!\right)_{\!\!,\mu}\!\!,
\end{equation}
hence
\begin{equation} \label{R3c}
	0=\delta S_{\rm m} = - \!\int\!T^{\mu\nu}g_{\mu\nu}\left(\!\frac{\delta\Omega}{\Omega}\!\right)\!\sqrt{-g}\,d^4x
	- 2\int\!G^\mu \left(\!\frac{\delta\Omega}{\Omega}\!\right)_{\!\!,\mu} \!\!\sqrt{-g}\,d^4x.
\end{equation}
Removing the full derivative, one obtains 
\begin{equation} \label{deltag}
	2(G^\mu)_{\,;\mu}={\rm Trace}[T^{\mu\nu}].
\end{equation}
This can be called ``the self-consistency condition''. Of cause it is hidden in the complete set of the field equations, but nevertheless, may be rather helpful and instructive. Note, that it contains not the Weyl covariant derivative, but the metric one.

\section{Description of particle production}

Particle production is a pure quantum process. This why there are enormous difficulties to describe it in the framework of any pure classical gravitational theory. Thee main problem is to take into account the back reaction of the quantum process (which is energetic) onto the spacetime geometry. Indeed, in order to solve correctly the quantum problem, one needs to impose some boundary conditions. At the same time,  In order to impose properly some boundary conditions, one needs to know  the spacetime structure, In order to know the spacetime structure, one has to solve the gravitational equations. In order to solve the gravitational equations, one needs to know the gravitating source, i.\,e., the energy-momentum tensor. But the energy-momentum tensor includes not only the classical matter fields and contributions from already created particles, it necessary includes the very process of their creation. Thus, we have the vicious circle.

One way to break this circle is to ignore completely the back reaction. This was done many than 50 years ago \cite{Parker69,GribMam70,ZeldStar72, ParFull73, HuFullPar73,FullParHu74,FullPar74, GribMamMostep76} . We have chosen another way: to describe the particle creation phenomenologically with full account for the back reaction. 

We will describe the already created particles hydrodynamically  by the perfect fluid. Lagrangian picture s not suitable for us because the number of trajectories is not constant, so we have to use the Eulerian picture. For usual hydrodynamics (respecting the particle number conservation law) the corresponding action integral was found by J. R. Ray \cite{Ray}. It reads as follows
\begin{equation} \label{R3}
	S_{\rm m}= -\!\int\!\varepsilon(X,n)\sqrt{-g}\,d^4x + \!\int\!\lambda_0(u_\mu u^\mu-1)\sqrt{-g}\,d^4x
+\!\!\int\!\lambda_1(n u^\mu)_{;\mu}\sqrt{-g}\,d^4x + \!\!\int\!\lambda_2 X_{,\mu}u^\mu\sqrt{-g}\,d^4x.
\end{equation}
The dynamical variables are $n(x)$ --- particle number density, $u^\sigma(x)$ ---  vector field, $X(x)$ ---  auxiliary variable. The Lagrange multipliers $\lambda$-s provide us with the constraints: $\lambda_0$ makes the four velocity out of vector fields, $\lambda_1$ insures the particle number conservation, $\lambda_1$ shows that $X(x)$ is constant along the trajectories (thus enumerating them), and $\varepsilon(X,n)$ is the matter energy density.

The modification allowing the creation of particles was made by V. A. Berezin \cite{Berezin87}. It consists simply in replacing the particle conservation law by some particle creation law, namely 
\begin{equation} \label{Phi2}
	(nu^\sigma)_{;\sigma}-\Phi(\rm inv)=0,
\end{equation}
where the ``creation'' function $\Phi(\rm inv)$ depends on the invariants constructed of the geometric and matter variables.

Since we are dealing with the Weyl geometry, the energy density $\varepsilon$ may depend also on new invariant $B$, absent in Riemannian geometry,
\begin{equation} \label{Phi2}
	B=A_\mu u^\mu.
\end{equation}

\section{Particle production rate} 

It was shown in  \cite{Parker69,GribMam70,ZeldStar72, ParFull73, HuFullPar73,FullParHu74,FullPar74, GribMamMostep76} that the main role in particle production (in one loop approximation) is played by the so called conformal anomaly. It consists of two parts, the local and non-local ones. It is just the non-local terms that are responsible for the particle creation, while the local terms describe the vacuum polarization by quantum fluctuations The latter consists of the quadratic combinations of the curvature tensor, Ricci tensor and curvature scalar. Note, that the appearance of quadratic curvature terms in gravitational Lagrangian was advocated by A. D. Sakharov in 1966 \cite{Sakharov66}

To understand how it could be done, let us investigate the behavior of creation law under the conformal transformation. Using the obvious relations
\begin{equation} \label{numu3c}
	n=\frac{\hat n}{\Omega^3}, \quad u^\mu=\frac{\hat u^\mu}{\Omega}, \quad \sqrt{-g}=\Omega^4 \sqrt{-\hat g},
\end{equation}
one has 
\begin{equation} \label{numu3}
	(n u^\sigma)_{;\sigma} =\frac{(n u^\sigma\sqrt{-g})_{,\sigma}}{\sqrt{-g}}
	=\frac{(\hat n\hat  u^\sigma\sqrt{-\hat g})_{,\sigma}}{\sqrt{-g}}.
\end{equation}

Thus $\Phi\sqrt{-g}$ is invariant under the conformal transformation. In Weyl gravity the quadratic curvature part is just the Weyl Lagrangian, ${\cal{L}_{\rm  W}}$ (may be with different coefficients). What else?

In result, we found two more combinations, both of them depends on the number density of the already created particles. They are
$n^{4/3}$ and $n(B+2\nabla_\mu u^\mu)$. 

Thus our creation law becomes 
\begin{equation} \label{PhiWeyl}
	\Phi= \alpha_1'R_{\mu\nu\lambda\sigma}R^{\mu\nu\lambda\sigma}
+\alpha_2'R_{\mu\nu}R^{\mu\nu}
+\alpha_3'R^2+ \alpha_4'F_{\mu\nu}F^{\mu\nu} 
+\gamma_1 n(B+2\nabla_\mu u^\mu) +\gamma_2 n^{4/3}.
\end{equation}

\section{Conclusions and Discussions}

We have found two brand new contributions to the particle creation law in the Weyl geometry. First of them initiates the dust matter, while the second one initiates the radiation. Note,however, that the signs o f the energy density may be negative. It is interesting that something like the bulk viscosity suddenly appeared. This term has no counterpart in the Riemanian geometry.

The Weyl gravity \cite{Weyl}, by definition, is invariant under the local conformal transformation. 

We are planning to apply the obtained results to the cosmology in the Weyl geometry.

\section*{Acknowledgments} 

{Authors are grateful to E. O. Babichev, Yu. N. Eroshenko, N. O. Nazarova and A. L. Smirnov for stimulating discussions.}


\begin{thebibliography}{99}
\bibitem{Sakharov66} Sakharov A D 1966 \emph{JETP} {\bf 22} 241
\bibitem{Vilenkin} Vilenkin A 1982 \emph{Phys. Lett. B}, {\bf 117} 25
\bibitem{Penrose} Penrose R 2014 \emph{Found. Phys.} {\bf 44} 873
\bibitem{Hooft} ’tHooft G 2015 \emph{arXiv} arxiv:1511.04427 [gr-qc]
\bibitem{Weyl} Weyl H 1918  \emph{Math. Zeit.} {\bf 2} 384
\bibitem{bde19} Berezin V, Dokuchaev V, Eroshenko Yu 2019  \emph{IJMPD} {\bf 28} 1941007  
\bibitem{bde20} Berezin V, Dokuchaev V, Eroshenko Yu 2020 \emph{IJMPA} {\bf 35} 2040002
\bibitem{bdes} Berezin V, Dokuchaev V, Eroshenko Yu, Smirnov A 2016 \emph{JCAP} {\bf 2016} 019.
\bibitem{bd23} Berezin V, Dokuchaev V 2023 \emph{Class. Quantum Grav.} {\bf 40} 015006
\bibitem{Parker69} Parker L 1969  \emph{Phys. Rev.} {\bf 183} 1057
\bibitem{GribMam70} Grib A A, Mamaev S G 1970 \emph{Sov. J. Nucl. Phys.} {\bf 10} 722
\bibitem{ZeldStar72} Zel'dovich Ya B, Starobinskii A A 1972 \emph{Sov. Phys. JETP} {\bf 34} 1159
\bibitem{ParFull73} Parker L and Fulling S A 1973 \emph{Phys. Rev. D} {\bf 7}  2357
\bibitem{HuFullPar73} Hu B L, Fulling S A, Parker L 1973 \emph{Phys. Rev. D} {\bf 8} 2377
\bibitem{FullParHu74} Fulling S A, Parker L, Hu B L 1974 \emph{ Phys. Rev. D} {\bf 10} 3905
\bibitem{FullPar74} Fulling S A, Parker L 1974 \emph{Ann. Phys.} {\bf 87} 176
\bibitem{GribMamMostep76} Grib A A, Mamaev S G, Mostepanenko V M 1976 \emph{Gen. Relativ. Gravit.} {\bf 7} 535
\bibitem{Ray} Ray J R 1972 \emph{J. Math. Phys.} {\bf 13} 1451
\bibitem{Berezin87} Berezin V A 1987 \emph{Intern. J. Mod. Phys. A} {\bf 2}  1591
\end{thebibliography}
\end{document}